\documentclass[sigconf]{acmart}

\usepackage{tikz}
\usepackage{bm}
\usepackage{graphicx}
\usepackage{placeins}
\usepackage{multirow}
\usepackage{booktabs}
\usepackage{array} 
\usetikzlibrary{arrows.meta, positioning, calc}
\AtBeginDocument{%
  }

\setcopyright{acmlicensed}
\copyrightyear{2025}
\acmYear{2025}
\acmDOI{XXXXXXX.XXXXXXX}
\acmConference[Conference acronym '25]{Make sure to enter the correct
  conference title from your rights confirmation email}{Oct 27--31,
  2025}{Woodstock, NY}
\acmISBN{978-1-4503-XXXX-X/2018/06}

\begin{document}
\settopmatter{printacmref=false} 

\title{Latent Space Consistency for Sparse-View CT Reconstruction}



\author{Duoyou Chen}
\email{chandery1112@gmail.com}
\orcid{1234-5678-9012}
\affiliation{%
  \institution{University of Science and Technology Beijing}
  \city{Beijing}
  \country{China}
}

\author{Yunqing Chen}
\email{Serein7z@gmail.com}
\affiliation{%
  \institution{University of Science and Technology Beijing}
  \city{Beijing}
  \country{China}
}

\author{Can Zhang}
\affiliation{%
  \institution{University of Science and Technology Beijing}
  \city{Beijing}
  \country{China}
}
\email{zhangcan66@163.com}

\author{Zhou Wang}
\affiliation{%
  \institution{University of Science and Technology Beijing}
  \city{Beijing}
  \country{China}
}

\author{Cheng Chen}
\authornote{Contact author}
\affiliation{%
  \institution{University of Science and Technology Beijing}
  \city{Beijing}
  \country{China}
}

\author{Ruoxiu Xiao}
\affiliation{%
  \institution{University of Science and Technology Beijing}
  \city{Beijing}
  \country{China}
}

\renewcommand{\shortauthors}{Trovato et al.}

\begin{abstract}
Computed Tomography (CT) is a widely utilized imaging modality in clinical settings. Using densely acquired rotational X-ray arrays, CT can capture 3D spatial features. However, it is confronted with challenged such as significant time consumption and high radiation exposure. CT reconstruction methods based on sparse-view X-ray images have garnered substantial attention from researchers as they present a means to mitigate costs and risks. In recent years, diffusion models, particularly the Latent Diffusion Model (LDM), have demonstrated promising potential in the domain of 3D CT reconstruction. Nonetheless, due to the substantial differences between the 2D latent representation of X-ray modalities and the 3D latent representation of CT modalities, the vanilla LDM is incapable of achieving effective alignment within the latent space. To address this issue, we propose the Consistent Latent Space Diffusion Model (CLS-DM), which incorporates cross-modal feature contrastive learning to efficiently extract latent 3D information from 2D X-ray images and achieve latent space alignment between modalities. Experimental results indicate that CLS-DM outperforms classical and state-of-the-art generative models in terms of standard voxel-level metrics (PSNR, SSIM) on the LIDC-IDRI and CTSpine1K datasets. This methodology not only aids in enhancing the effectiveness and economic viability of sparse X-ray reconstructed CT but can also be generalized to other cross-modal transformation tasks, such as text-to-image synthesis. We have made our code publicly available at https://anonymous.4open.science/r/CLS-DM-50D6/  to facilitate further research and applications in other domains.

\end{abstract}




\maketitle

\section{Introduction}
\label{sec:intro}
With the continuous advancement of medical imaging technology, Computed Tomography (CT) has emerged as a vital instrument in clinical diagnosis. Traditional CT imaging techniques involve the acquisition of data through multiple angles and planes of X-ray images, which are subsequently utilized to generate 3D reconstructed images. However, this conventional approach poses significant economic costs in terms of time and radiation risks. Consequently, the application of sparse-view X-ray images for CT reconstruction can effectively reduce both the scanning costs and the health hazards posed to patients. Nevertheless, the primary challenge lies in how to effectively extract the implicit three-dimensional information contained within the X-ray images.

\begin{figure}
    \centering
    \includegraphics[width=\linewidth]{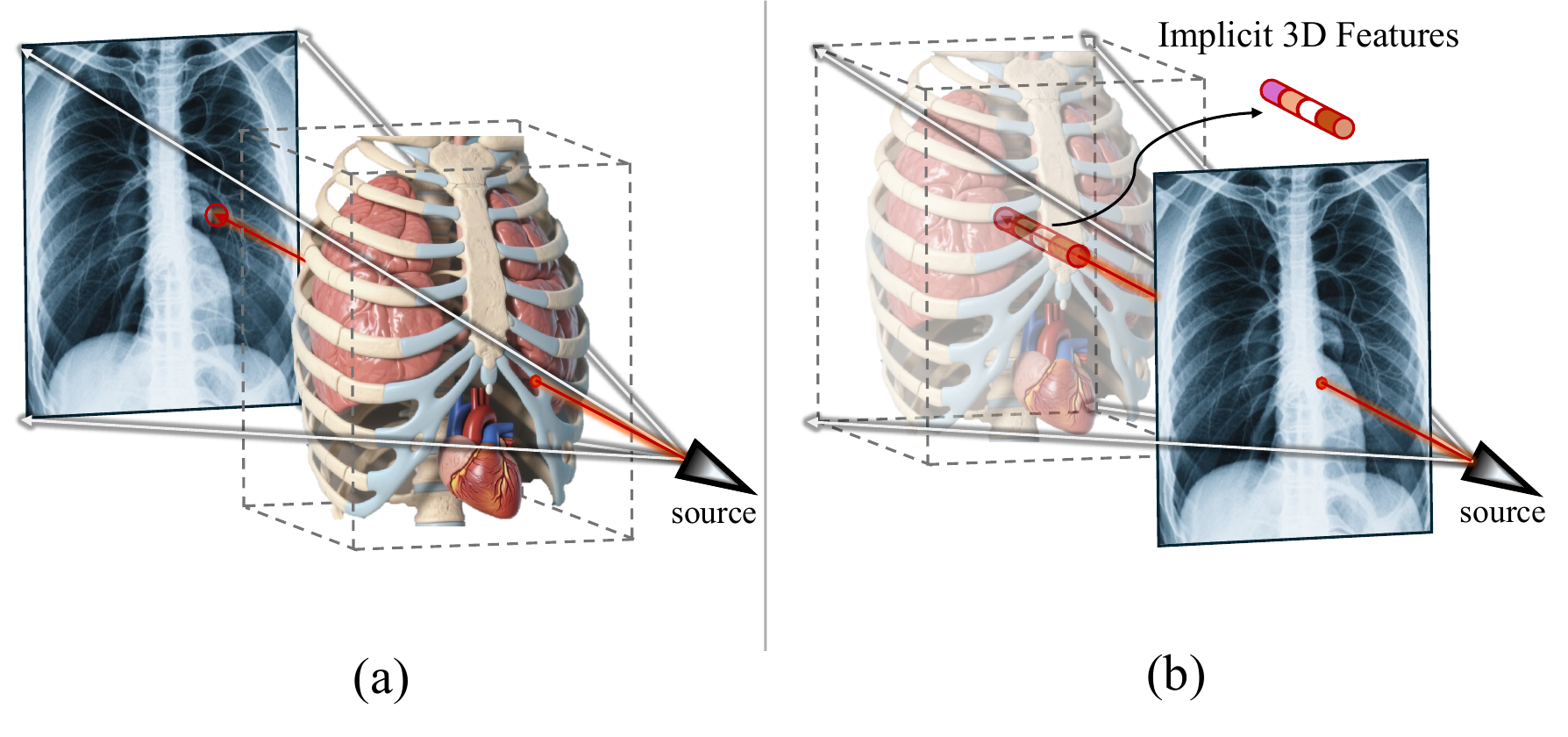}
    \caption{The diagram illustrates two logically opposite imaging processes. (a) depicts the process by which X-rays emitted from a light source traverse through an imaging entity, resulting in the projection of a coronal X-ray image. Conversely, (b) illustrates the logically reversed process—extracting three-dimensional information of the imaging entity from the provided coronal X-ray image.}
    \label{fig:xray}
\end{figure}

Variational Autoencoders (VAE), Generative Adversarial Networks (GAN), and Denoising Diffusion Probabilistic Models (DDPM) have found extensive applications in the task of CT reconstruction utilizing sparse-view X-ray images \cite{b1,b2}. Specifically, these generative models employ different conditional encoders to extract the implicit 3D spatial information from sparse X-ray images and utilize probabilistic distribution models to learn the mapping relationship between these images and their corresponding CT outputs, thus enabling the reconstruction of 3D CT images from sparse X-ray inputs. Notably, significant research has been conducted on the application of Generative Adversarial Networks (GANs), demonstrating their efficacy in this domain \cite{x2ct}. However, a severe limitation is posed by the inherent instability during GAN training and the phenomenon of mode collapse\cite{gbd,zero}, which complicates further improvements in performance.

Diffusion models employ a Markov chain to iteratively introduce noise\cite{ddpm}, followed by a reverse process to denoise, thereby addressing the divergence issue between the latent space of VAE\cite{fdp} and the standard Gaussian distribution. Additionally, this approach mitigates the mode collapse problem commonly associated with GANs\cite{cdp}. However, a significant drawback of diffusion models is the substantial increase in spatiotemporal costs required for training. The emergence of Latent Diffusion Models (LDM)\cite{m5} has allowed for the compression of images from pixel space into latent space, effectively maintaining training efficacy while markedly reducing the costs associated with training and inference, thus enabling the practical use of diffusion models in 3D contexts. Previous research has reported numerous instances of utilizing LDM for X2CT reconstruction, demonstrating notable advantages over baseline studies reliant on GANs \cite{b4}. Nonetheless, a primary challenge remains: the conditional feature vectors within the latent space often fail to align effectively with the latent space itself when guiding the diffusion process. This misalignment increases the learning pressure during the diffusion process and can hinder convergence.

To address this issue, we propose CLS-DM, a contrastive learning-guided Latent Diffusion Model tailored for application in 3D CT sparse view reconstruction. Our innovative approach includes a pre-training phase for a conditional encoder, wherein contrastive learning is integrated to align the X-ray feature space with the latent space. Furthermore, to mitigate the potential degradation of feature representation capability in the conditional encoder due to contrastive learning, we incorporate an autoregressive process to guide the training of the conditional encoder, thereby enhancing its capacity for extracting 3D information. Recognizing that capturing X-ray images from unconventional angles incurs additional costs, our method restricts the selection of X-ray views to only a subset of those in the sagittal and coronal planes. This strategic choice not only leads to higher-quality reconstructions but also presents a more feasible solution for clinical practice.

In summary, our work presents the following contributions:
\begin{itemize}
    \item \textbf{Introduction of the Contrastive Learning Module}: Inspired by LDM, we innovatively integrate a contrastive learning module, which effectively aligns the X-ray feature vectors input during the diffusion process with the latent space of the diffusion itself.
    \item \textbf{Optimization of the Conditional Encoder Guided by Autoregression}: We employ an autoregressive approach to optimize the conditional encoder. This strategy safeguards the feature representation capability of the conditional encoder against degradation.
    \item \textbf{Efficiency and Real-time Performance}: Although our model incorporates a contrastive learning phase compared to existing single-phase and two-phase generative models, the inference process does not significantly increase computational complexity. 
\end{itemize}

\section{Related Works}
\label{sec:formatting}

\subsection{Sparse-view Computed Tomography Reconstruction}
Sparse-view CT reconstruction is generally categorized into extremely few views reconstruction (EFVR), which involves the use of a single or biplanar view \cite{a6,a7,a8,a9}; few views reconstruction (FVR), employing fewer than ten views \cite{a5,a10}; and normal sparse-view reconstruction (NSVR), which utilizes dozens to hundreds of views \cite{a1,a2,a3,a4}. Our work falls under the framework of EFVR. Prior investigations have explored this domain using both supervised and generative models: 

Supervised models primarily employ CNNs or INRs. For instance, Sun et al. \cite{a8} utilized an INR structure for CT reconstruction with single and biplanar X-rays, demonstrating applicability in low-resource radiotherapy with less than 1\% error in radiation dose calculations compared to standard CT scans. Lin et al. (2023) \cite{a12} meticulously structured their approach by formulating CT reconstruction as a continuous intensity field, leading to the development of the DIF-Net framework, which extracts and aggregates view-specific features for precise intensity regression. However, many of these supervised model studies utilize mean squared error (MSE) or mean absolute error (MAE) losses, resulting in overly smooth outcomes. To address this issue, several researchers have attempted to leverage generative models. Ying et al. \cite{a6} and Cafaro et al. \cite{a11} employed GANs to construct their methods. More recent studies have adopted diffusion models, which can produce outputs with richer and more accurate details compared to GANs; however, their high computational costs render 3D reconstruction tasks challenging. Chung et al. \cite{a13} and Lee et al. \cite{a14} attempted to process each 2D slice in 3D CT using 2D diffusion models, which sacrifices consistency and relationships between slices \cite{a5}. Sun et al. \cite{a5} introduced the LDM framework to compress CT from voxel space to latent space, significantly reducing the computational costs of the diffusion process, albeit at the cost of introducing inconsistencies between the latent space and X-ray feature space. We have designed the CLS-DM, which ingeniously integrates contrastive learning to align the two spaces prior to the diffusion process.

\begin{figure*}[htb!]
    \centering
    \includegraphics[width=\linewidth]{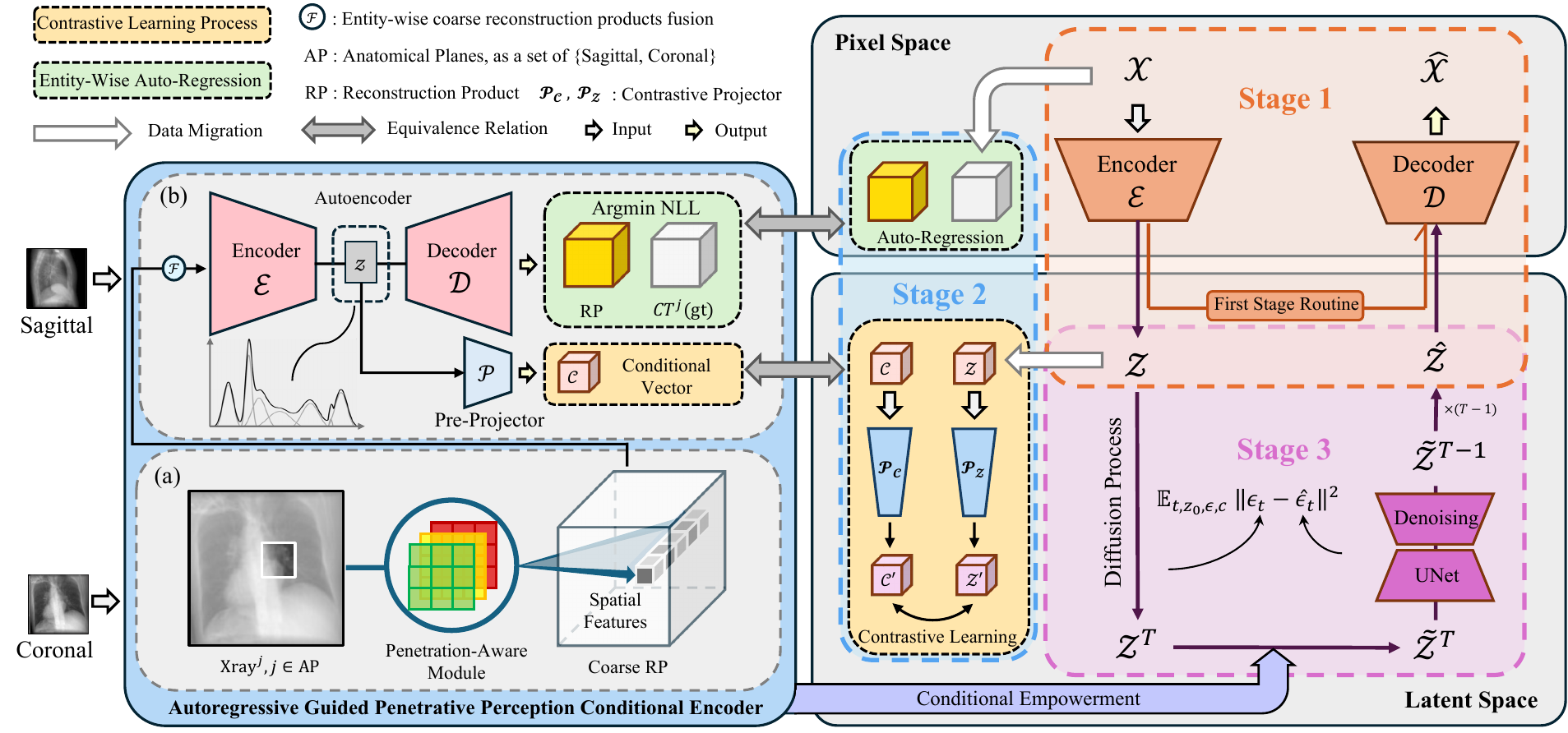}
    \caption{The complete workflow of the CLS-DM algorithm, encompassing both the conditional encoder (left portion) and the latent diffusion process (right portion). The conditional encoder is divided into two components: (a) and (b). (a) depicts the operational principles and functionalities of the perceptual compression module, while(b) outlines the workflow of the autoencoder and its relationship with downstream tasks. The latent diffusion process is delineated into three stages, represented by the colors orange, blue, and purple, corresponding to perceptual feature compression, latent space alignment, and the diffusion stage within our algorithm. It is noteworthy that the diagram standardizes the color coding for each trainable module across the stages, indicating where each module is trained and where it is frozen. For Stage One, the trainable modules Encoder and Decoder are marked in orange; for Stage Two, the trainable conditional encoder (left portion) and two projection heads ($\mathcal{P}_\mathcal{C}$ and $\mathcal{P}_\mathcal{Z}$) are consistently marked in blue; and for Stage Three, the trainable module UNet is indicated in purple.
    }
    \label{fig:method}
\end{figure*}

\subsection{Contrastive Learning Framework}
Contrastive learning facilitates the acquisition of discriminative feature representations by minimizing the distance between positive sample features while maximizing the distance among negative samples, serving as an effective tool for cross-modal alignment. Prominent frameworks in this domain include SimCLR \cite{ba}, MoCo \cite{bb}, and BYOL \cite{bc}. The SimCLR framework proposed by Chen et al.\cite{ba} generates different modal views of the same entity through data augmentation, employing the InfoNCE loss to maximize the similarity of positive sample pairs. This approach is advantageous due to its simplicity and efficiency, requiring no negative sample queue.Conversely, the MoCo framework introduced by He et al.\cite{bb} incorporates a momentum encoder and a queue of negative samples, which enhances the quality of feature representations but increases training complexity. Meanwhile, BYOL, designed by Grill et al.\cite{bc}, utilizes a self-supervised approach that obviates reliance on negative samples; however, it is susceptible to feature collapse.

In the context of medical cross-modal reconstruction, the strength of SimCLR lies in its compatibility with diverse modal data augmentation strategies\cite{bd}. Given that CT and X-ray modalities exhibit distinct imaging characteristics and noise distributions, the flexible augmentation mechanism of SimCLR can better accommodate these disparities without the need for intricate queue management. In contrast to alternative frameworks, SimCLR explicitly optimizes feature discrimination through negative sample comparison, thereby ensuring effective alignment between compressed CT features and encoded X-ray features in the latent space, as well as the learning of patterns during the diffusion phase\cite{be}. Consequently, our CLS-DM adopts the SimCLR contrastive learning framework to enhance the similarity of positive samples while promoting the distinction of different entities, thereby reducing the distance between the X-ray feature space and the latent space.

\subsection{Multi-Stage Image Synthesis}
A considerable body of research \cite{m1,m2,m3,m4,m5} has employed multi-stage synthesis to progressively refine image details through hierarchical processing, establishing this approach as a standard paradigm for high-resolution image generation. Variational Quantized Autoencoders (VQVAEs) \cite{m3} construct a discrete latent space via vector quantization; VQGAN \cite{m2} enhances generation quality by integrating Generative Adversarial Networks (GANs); and Latent Diffusion Models (LDM) \cite{m5} balances efficiency and accuracy through a two-stage design consisting of perceptual compression and diffusion. 

However, existing two-stage models lack explicit alignment of multi-modal features during the conditional encoding phase, leading the subsequent diffusion process to overly rely on noise priors rather than structured conditional information. In response to this limitation, the CLS-DM proposed in this paper innovatively adopts a three-stage training framework: the first stage compresses CT data from voxel space into latent space; the second stage aligns conditional features with latent space under autoregressive reconstruction constraints; and the third stage performs diffusion generation within the aligned latent space. This design builds upon and enhances the temporal constraints of LDM, clearly delineating the functionalities and impacts of each module. Such improvements facilitate more precise identification of problem modules during the hyperparameter tuning phase and enhance the efficiency of training fine-tuning in diverse task contexts.

\section{Methods}
LDM effectively reduces the computational demands associated with high-resolution image synthesis by compressing images from voxel space into latent space. However, this approach introduces a subsequent challenge wherein the conditional encoder struggles to align the distribution within the latent space during the diffusion process, resulting in increased learning pressure and suboptimal training outcomes. Moreover, medical images typically exhibit specific anatomical structures and pathological features, contributing to a latent space that is relatively continuous and entangled\cite{aa1}. Consequently, the issue of latent space alignment becomes particularly pronounced in the context of reconstructing medical images.

We propose a strategy that aligns the latent space formed by perceptual compression with the space of the conditional encoder prior to the diffusion process, thereby enhancing the training of the diffusion process from both efficiency and accuracy perspectives(See Figure \ref{fig:method}). However, indiscriminate alignment may compromise the feature perception capability of the conditional encoder, potentially leading to a loss of important features. We introduce a contrastive learning module between the perceptual compression and the diffusion process. This module not only facilitates the alignment of spaces but also employs an autoregressive approach to guide the optimization of the conditional encoder.

\subsection{3D Conditional Latent Diffusion Model}
\subsubsection{Perceptual Feature Compression}

An effective perceptual feature compression process necessitates the provision of a latent space capable of representing the high-dimensional semantic features of CT images while attenuating spatial redundant information. We employ a GAN\cite{aa2} framework that incorporates a generator-discriminator architecture. The generator component utilizes a Variational Autoencoder (VAE)\cite{aa3}, of which each feature within the latent space is represented by a Gaussian distribution. This representation is designed to optimize the fitting of specific pathological and anatomical characteristics inherent to medical images.

In the development of our model, we utilize a 3D VAE that comprises an encoder function \( f_E(\cdot) \) and a decoder function \( f_D(\cdot) \). The encoder is designed to map the input CT image \( X^i \in \mathbb{R}^{1 \times d \times h \times w} \) into the latent space, resulting in an output represented as \( Z^i \in \mathbb{R}^{2 \times c \times d' \times h' \times w'} \), where it is assumed that \( d,h,w > c d' h' w' \).

The variable \( Z^i \) is partitioned into two distinct components along the channel dimension: \( \mu^i \in \mathbb{R}^{c \times d' \times h' \times w'} \), which indicates the feature mean, and \( \ln\widetilde{\varepsilon}^i \in \mathbb{R}^{c \times d' \times h' \times w'} \), which signifies the logarithm of the variance matrix of the latent variables. During each instance of forward propagation, the reparameterization trick\cite{aa3} is utilized to sample the latent variables \( Z_0^i \in \mathbb{R}^{c \times d' \times h' \times w'} \) from the Gaussian distribution characterized by the feature mean and variance matrix. These sampled latent variables subsequently serve as input to the decoder network. The process of reparameterization can be mathematically articulated as Equation \ref{equl:repara}, with \( \varepsilon \sim \mathcal{N}(0, I) \).

\begin{equation}
Z_0^i=\mu^i+\mathrm{e}^{{ln\widetilde{\varepsilon}}^i}\times\varepsilon
\label{equl:repara}
\end{equation}

The decoder subsequently maps the sampled \( Z_0^i \) back into the voxel space to facilitate reconstruction, with the output denoted as \( X^i \in \mathbb{R}^{1 \times d \times h \times w} \). To enhance the determinism of the reconstruction results rather than their diversity, during the subsequent stages, including inference, we directly utilize the mean \( \mu^i \) in place of the sampled outcomes. In our experiments, \( d \times h \times w = 128^3 \) and \( d' \times h' \times w' = 16^3 \), with \( c = 4 \).

In the optimization process of GAN, the generator and discriminator utilize loss functions denoted as \( \mathcal{L}_G \) and \( \mathcal{L}_D \), respectively. The loss function for the generator, \( \mathcal{L}_G \), comprises the autoregressive Negative Log-Likelihood Loss (NLL Loss) \( \mathcal{L}_{nll} \) (as presented in Equation \ref{equl:nll}), the latent space Kullback-Leibler Divergence Loss \( \mathcal{L}_{kl} \) (as detailed in Equation \ref{equl:kl}), and the perceptual loss \( \mathcal{L}_P \) based on the discriminator. Consequently, \( \mathcal{L}_G \) can be succinctly expressed as indicated in Equation \ref{equl:lg}.

\begin{equation}
    \mathcal{L}_{nll}(\sigma)=\frac{\left(X^i-{\hat{X}}^i\right)^2}{2\sigma^2}+\log{\sigma}+C
\label{equl:nll}
\end{equation}
 \( \sigma \) represents the uncertainty, which is a learnable parameter, while \( C \) is defined as a constant.

\begin{equation}
    \mathcal{L}_{kl}=\frac{{\mu^i}^2+\mathrm{e}^{{ln\widetilde{\varepsilon}}^i}-1-{ln\widetilde{\varepsilon}}^i}{2}
\label{equl:kl}
\end{equation}

\begin{equation}
    \mathcal{L}_G=\lambda_1\mathcal{L}_{nll}(\sigma)+\lambda_2\mathcal{L}_{kl}+\lambda_3\mathcal{L}_{P}
\label{equl:lg}
\end{equation}
where \( \lambda_1 \), \( \lambda_2 \), and \( \lambda_3 \) are coefficients that regulate the contribution of each loss function within the overall loss formulation. The loss function \( \mathcal{L}_D \) employed is derived from the Hinge Discriminator Loss, and its formula is expressed as Equation \ref{equl:ld}.

\begin{equation}
    \begin{aligned}
    \mathcal{L}_D&=\frac{1}{2}\mathbb{E}_{x\sim p_x}\left[\max{\left(0,1-D\left(x\right)\right)}\right] \\
    &+\frac{1}{2}\mathbb{E}_{z\sim p_z}\left[\max{\left(0,1+D\left(G\left(z\right)\right)\right)}\right]
    \end{aligned}
\label{equl:ld}
\end{equation}
with \( p_x \) denotes the distribution of CT images, while \( p_z \) represents the distribution in the latent space.

\subsubsection{Conditional Diffusion Process}

During the diffusion process, we employ a time-conditional UNet \cite{aa4,aa5} to learn the noise \( \epsilon \) in the inverse Markov chain. Specifically, the output of the contrastively learned conditional encoder \( \tau_\theta(y) \) (as detailed in \ref{sec:cl}\& \ref{sec:ce}) is concatenated with the latent variable \( z_t \) at time \( t \) along the channel dimension, and this combined representation serves as the guiding condition for the UNet in order to obtain the noise estimate \( \hat{\epsilon}_t \). This process is typically denoted as \( \hat{\epsilon}_t = \epsilon_\theta(z_t, t, \tau_\theta(y)) \), where \( \epsilon_\theta \) indicates the mapping function of the UNet. Consequently, the optimization objective for the conditional diffusion process can be expressed as Equation \ref{equl:Lldm}

\begin{equation}
\mathcal{L}_{LDM} = \mathbb{E}_{t, z_0, \epsilon, c} [\epsilon_t - \epsilon_t^2]
\label{equl:Lldm}
\end{equation}

In the training of the conditional diffusion process, we randomly generate a set of timestamps \( \dot{t} = \{t_1, t_2, \ldots, t_s\} \) (where \( t_i, s \leq T \); \( T \) represents the total length of the Markov chain) for the optimization of the UNet model. In our experiments, we set \( T = 1000 \) for model training. During the inference phase, we employed the DPM-Solver \cite{aa8} method for sampling, which is capable of restoring the sampling effects of the Markov chain with minimal steps, all without the need for fine-tuning the model.
\begin{figure}[!t]
\centering
\includegraphics[width=\linewidth]{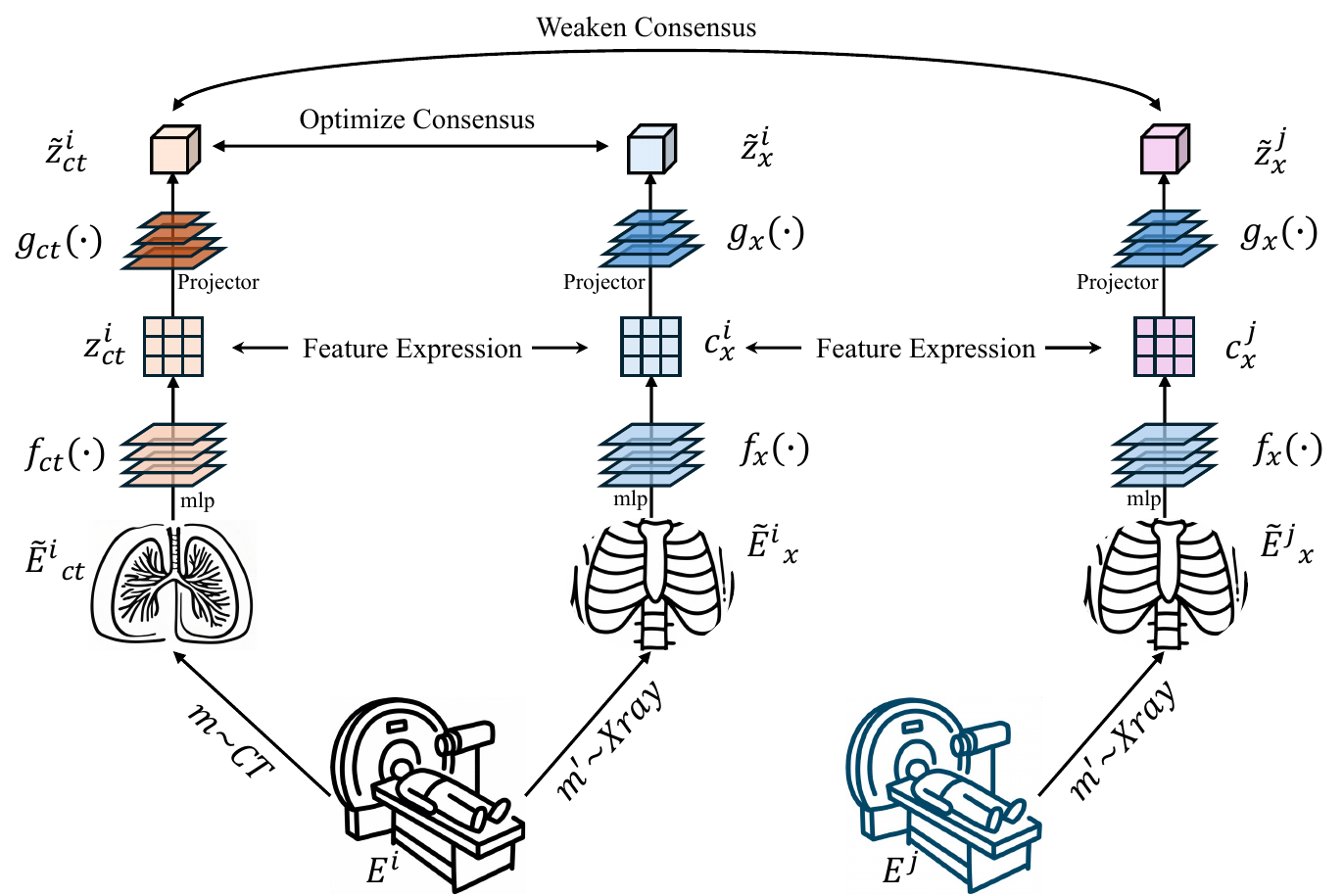}
\caption{The modality consistency optimization process based on the contrastive learning framework. In this process, feature representation learning is conducted for CT and X-ray images pertaining to the same entity. The optimization process aims to enhance the consistency among latent variables corresponding to the same entity while improving the distinguishability between latent variables of different entities (as demonstrated in the diagram by \( \widetilde{z}_x^i \) and \( \widetilde{z}_x^j \)). }
\label{fig:cl}
\end{figure}

\begin{table*}[htb!]
\centering
\caption{A quantitative comparison of different methods is presented under the LIDC-IDRI and CTSpine1K datasets. The optimal values are marked in bold, while the suboptimal values are highlighted with underlining.}
\begin{tabular}{l c|cc|cc|cc|cc|cc}
\toprule
Method & \multirow{2}{*}{X-Type} &\multicolumn{2}{c|}{X2CT-CNN [56]} & \multicolumn{2}{c|}{X2CT-GAN [35]} & \multicolumn{2}{c|}{INRR3CT [43]} & \multicolumn{2}{c|}{LDM [15]} & \multicolumn{2}{c}{CLS-DM(ours)} \\
\cline{3-12}
Scene & & PSNR↑ & SSIM↑ & PSNR↑ & SSIM↑ & PSNR↑ & SSIM↑ & PSNR↑ & SSIM↑ & PSNR↑ & SSIM↑  \\
\midrule
\multirow{3}{*}{LIDC-IDRI} & frontal & \underline{23.01} & 0.5743 & 21.88 & 0.5121 & 21.83 & 0.5614 & 20.65 & \underline{0.5864} & \textbf{23.02} & \textbf{0.5931}  \\
 & lateral & \textbf{25.71} & \underline{0.6481} & 24.08 & 0.5823 & 23.84 & 0.6266 & 24.73 & 0.6473 & \underline{24.96} & \textbf{0.6585}  \\
 & 2-view & \textbf{27.49} & \underline{0.7010} & 24.78 & 0.6322 & 24.91 & 0.6852 & 24.94 & 0.6515 & \underline{27.36} & \textbf{0.7058}  \\
\hline
\multirow{3}{*}{CTSpine1K} & frontal & 19.19 & 0.4909 & 18.71 & 0.4545 & 19.60 & 0.4177 & \underline{22.83} & \underline{0.5309} & \textbf{24.48} & \textbf{0.5526}  \\
 & lateral & 21.11 & 0.5259 & 19.18 & 0.4713 & 20.33 & 0.4834 & \underline{23.53} & \underline{0.5624} & \textbf{25.15} & \textbf{0.5826}  \\
 & 2-view & 20.33 & 0.5397 & 19.54 & 0.4900 & 20.49 & 0.5061 & \underline{24.79} & \underline{0.5992} & \textbf{26.37} & \textbf{0.6186}  \\
\midrule
Average & & 22.81 & 0.5800 & 21.36 & 0.5237 & 21.83 & 0.5467 & \underline{23.58} & \underline{0.5963} & \textbf{25.22} & \textbf{0.6185} \\
\bottomrule
\end{tabular}
\label{tab:comparison}
\end{table*}
\subsection{ Contrastive Learning Module}
\label{sec:cl}

\begin{figure}
    \centering
    \includegraphics[width=\linewidth]{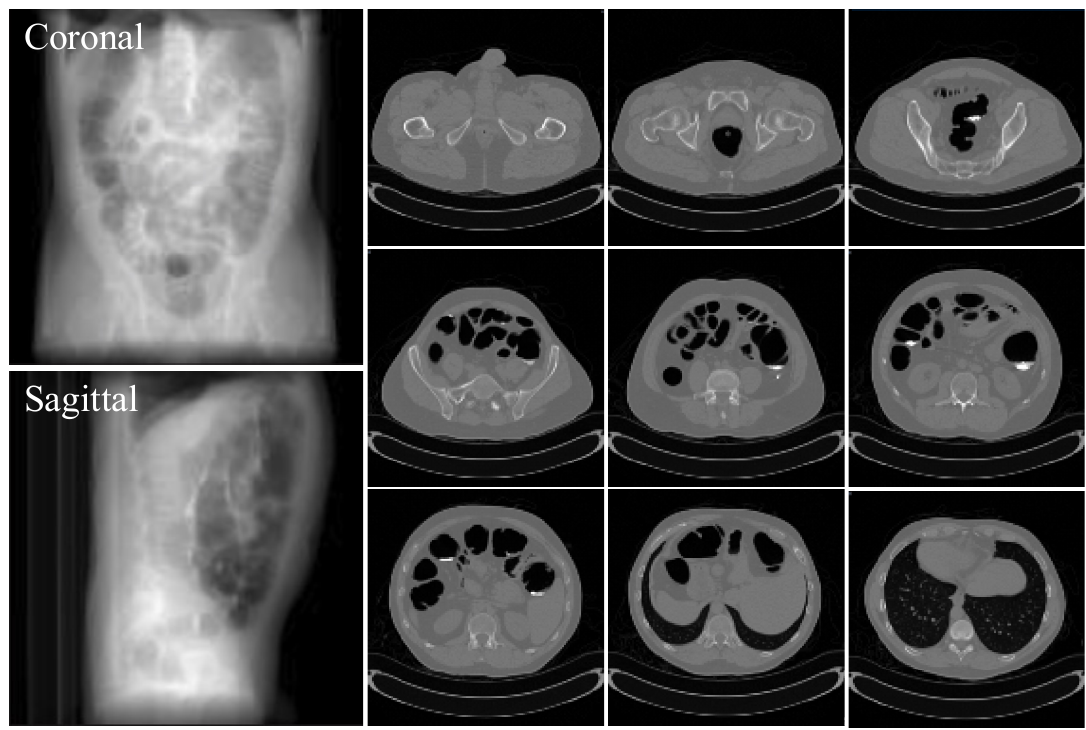}
    \caption{An example of Digital Reconstructed Radiography (DRR) algorithm outputs for coronal and sagittal plane X-ray and CT slices on dataset CTSpine1K.}
    \label{fig:data_example}
\end{figure}

To align the output space of the conditional encoder with the latent space, we introduce a contrastive learning module based on the SimCLR \cite{aa6} framework. Unlike the native application scenarios of the framework, our contrastive learning requires comparison across two distinct modalities: CT and X-ray images (see Fig. \ref{fig:cl}). For CT images \( \widetilde{E}_{ct}^i \) and X-ray images \( \widetilde{E}_{\mathcal{x}}^i \) that represent the same entity, we utilize modality-specific encoders \( f_{ct}(\cdot) \) and \( f_x(\cdot) \) to derive the feature representations \( z_{ct}^i \) and \( c_x^i \) for subsequent tasks.

In order to enhance the relational coherence between different modalities while simultaneously preventing the contrastive process from impairing the feature representation of the encoders, we constructed a contrastive learning framework that employs modality-specific projection heads \( g_{ct}(\cdot) \) and \( g_x(\cdot) \) to map feature space vectors into the latent space, resulting in \( \widetilde{z}_{ct}^i \) and \( \widetilde{z}_x^i \). 

\begin{equation}
    \mathcal{L}_{i,j}=\mathbb{E}_{i,j}-log\frac{exp\left(Sim\left(z_{ct}^i,z_x^i\right)/\mathcal{T}\right)}{\sum_{z_x^j}ex p\left(Sim\left(z_{ct}^i,z_x^j\right)/\mathcal{T}\right)}
\label{equl:infonce}
\end{equation}

During the optimization of the contrastive learning process, we aim to achieve a higher consistency for latent variables \( \widetilde{z}_{ct}^i \) and \( \widetilde{z}_x^i \) that originate from the same entity, while ensuring a greater degree of separation for latent variables \( \widetilde{z}_{ct}^i \) and \( \widetilde{z}_x^j \) that stem from different entities. We utilize the consistency evaluation metric defined as \( \text{Sim}(u, v) = \frac{u^T v}{\|u\| \|v\|} \) (cosine similarity), and employ the temperature-scaled InfoNCE Loss as the optimization objective \cite{aa7}(See Equation\ref{equl:infonce}), where $\mathcal{T}$ denotes a temperature parameter that scales the similarity scores. 

In the specific implementation, \( f_{ct}(\cdot) \) represents the encoder used in perceptual image compression, which is responsible for mapping the original CT images into the latent space, while \( f_x(\cdot) \) denotes the conditional encoder (as detailed in \ref{sec:ce}), tasked with extracting features from multiple X-ray images taken at different angles and integrating them into a conditional vector. Additionally, the positive and negative samples \( i \) and \( j \) in the loss function are sourced from the same mini-batch. This process enables the conditional encoder to indirectly learn the feature consistency between the two modalities and the consistency within the latent space while learning the regularization of latent variable coherence through contrastive learning.
\begin{figure*}
    \centering
    \includegraphics[width=\linewidth]{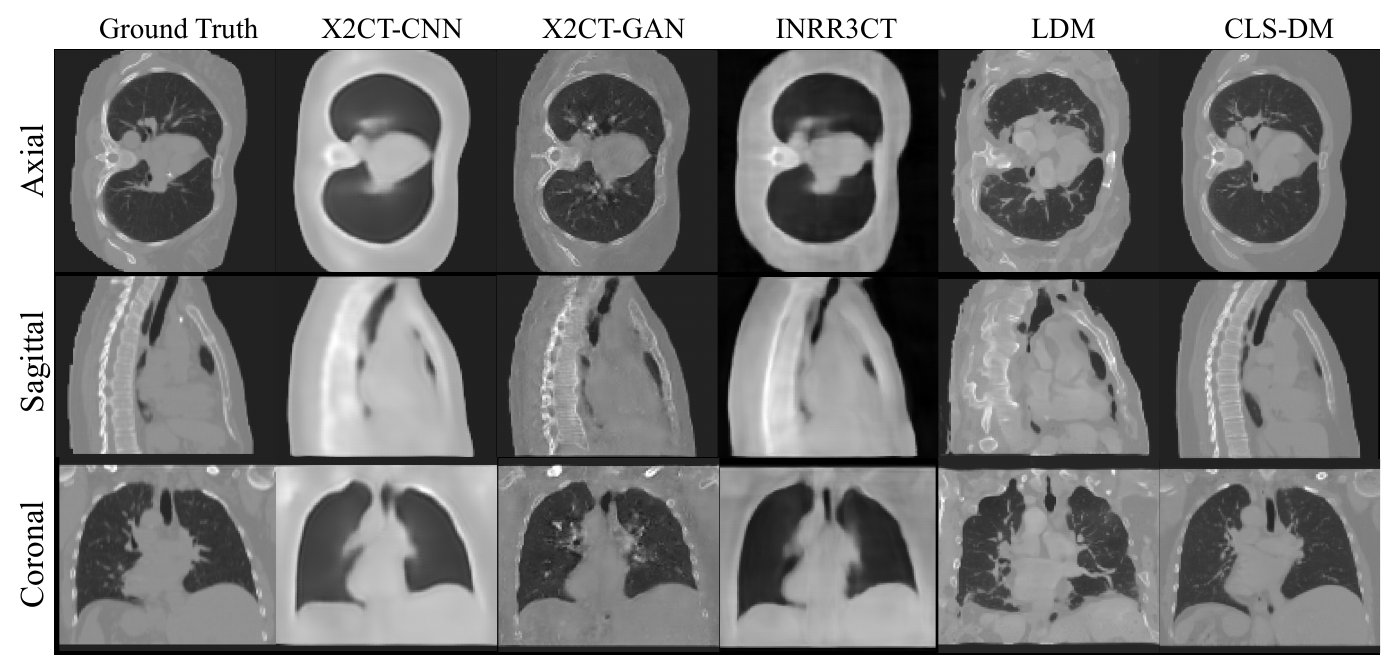}
    \caption{Visual comparison between the results of various models and the Ground Truth under the LIDC-IDRI dataset}
    \label{fig:lidc}
\end{figure*}
\subsection{Autoregressive-Guided Penetration-Aware Conditional Encoder}
\label{sec:ce}
A significant challenge arises in the conversion of two-dimensional X-ray images into latent variables that exhibit a shape analogous to that of perceptually compressed products derived from computed tomography (CT). Instead of merely performing straightforward replication and stacking of the X-ray images, we employ the X-ray Penetration-Aware Coarse Reconstruction Module (PAM) to transform the X-ray images into 3D coarse reconstruction outputs in voxel space (see Figure \ref{fig:method}-a).  Let the set of several X-ray images belonging to the same entity as \( Y^i = \{ Y_k^i \in \mathbb{R}^{1 \times h \times w} \ | \ k \in [1, K], K=1, 2 \} \). For each image \( Y^i \), the penetration-aware module \( \mathcal{P}(\cdot) \) initially extends the depth dimension by one unit, resulting in \( \widetilde{Y}_k^i \in \mathbb{R}^{1 \times 1 \times h \times w} \). Subsequently, a network architecture comprising several depth-wise transpose convolution layers maps this representation to a coarse reconstruction product \( \dot{Y}_k^i \in \mathbb{R}^{1 \times d \times h \times w} \), aligned in depth with the CT images.

Prior to conducting feature compression, the K coarse reconstruction products are fused together. We posit that the K X-ray images are captured within a negligible time interval, implying that there is no data offset caused by patient movement or other uncontrollable factors. These images are directly mapped into the same coronal space and averaged on a voxel-by-voxel basis to yield \( \dot{Y}^i \). Following this, perceptual feature compression is employed to condense \( \dot{Y}^i \) into a conditional variable \( c^i \in \mathbb{R}^{c' \times d' \times h' \times w'} \) aligned with the latent variables. To further mitigate the degradation of the conditional encoder's feature perceptual capability caused by the contrastive process, we incorporate an autoregressive mechanism to guide the training of the conditional encoder. Given the structure of the latent space within the latent diffusion model and the requirements for autoregressive guidance, architectures such as UNet and traditional autoencoders (AEs) present viable options. In our algorithm, the conditional encoder adopts a traditional AE framework (see Figure \ref{fig:method}-b), as the skip connections characteristic of UNet potentially bypass the feature representations in latent space, thereby diminishing the effectiveness of the autoregressive process. Specifically, we symmetrically introduce upsampling after feature compression, reconstructing the conditional variable \( c^i \) into a conditional reconstruction product \( \hat{Y}^i \in \mathbb{R}^{1 \times d \times h \times w} \) that is aligned with the CT shape. At this juncture, the autoregressive optimization process employs \( \hat{Y}^i \) in conjunction with the CT images \( X^i \) corresponding to the same entity, ensuring the accurate preservation of the conditional features. The optimization objective for this process, denoted as \( \mathcal{L}_{cond} \), can be expressed mathematically as follows:
\begin{equation}
\mathcal{L}_{cond} = \lambda_1 \mathcal{L}_{nll}(\sigma') + \lambda_2 \mathcal{L}_{clip}
\label{equl:lcond}
\end{equation}
Here, \( \mathcal{L}_{nll}(\sigma') \) represents the negative log-likelihood (NLL) loss associated with the autoregressive process, while \( \sigma' \) denotes the uncertainty. The term \( \mathcal{L}_{clip} \) signifies the loss incurred from contrastive learning (as detailed in \ref{sec:cl}). Furthermore, \( \widetilde{\lambda}_1 \) and \( \widetilde{\lambda}_2 \) are calibration parameters that determine the weight of each respective loss component.

\section{Experiments}
\subsection{Datasets}
We utilized the colonography subset from the publicly available datasets LIDC-IDRI\cite{lidc} and CTSpine1K\cite{ctspine}. The The LIDC-IDRI dataset includes a total of 1,018 diagnostic and lung cancer screening CT scans of the thorax. This dataset has been segmented into training, validation, and testing sets with a distribution of 716, 102, and 200 scans, respectively. The CTSpine1K dataset, which is specifically tailored for colonography, comprises 784 thoracic CT images. We have similarly divided this dataset into training, validation, and testing subsets containing 550, 78, and 156 images, respectively.

For the CT scans in the LIDC dataset, we extracted values within the range of [0, 2500], while for the CT scans in the CTSpine1K dataset, we retained values within a range of [-1024, 1476]. Prior to training, all CT images were normalized to a range of [-1, 1] and resized to a cubic dimension of \( 128^3 \).

Simultaneously, we employed the Digital Reconstructed Radiograph (DRR) algorithm to generate X-ray images at two angles, specifically 0° (Frontal) and 90° (Lateral), from each CT image(See Fig. \ref{fig:data_example}), effectively simulating the conditions of realistic X-ray imaging. The source-to-detector distance used in this process was set at 1020 mm, and the dimensions of the X-ray images were specified as 512 mm in height and width, with a pixel scale of 1 mm. Consistent with the treatment of the CT images, all X-ray images were also normalized to a range of [-1, 1] and resized to square images of \( 128^2 \).

\subsection{Metrics}
We employed reconstruction metrics, specifically Peak Signal-to-Noise Ratio (PSNR) and Structural Similarity Index Measure (SSIM), to evaluate the quality of the CT reconstructions. For both the LIDC and CTSpine1K datasets, we utilized 12-bit precision, applying Hounsfield unit value ranges of [0, 4095] for the LIDC dataset and [-1024, 3071] for the CTSpine1K dataset to compute these metrics.
For the 3D reconstruction data, we implemented a layer-by-layer calculation of the 2D metrics, averaging the results to obtain the final values. The PSNR metric ranges from 0 to 100 dB, with higher values indicating closer resemblance to the original image. Conversely, SSIM values range from 0 to 1, where values approaching 1 signify a higher degree of similarity between the images.

\subsection{Environment and Hyperparameters}
\subsubsection{CLS-DM}
The model was constructed using PyTorch and trained on an RTX 4090 24GB GPU for both training and inference. The shape of the latent space resulting from the compression phase of the perceptual characteristic is \( {16}^3 \). The number of channels in the latent variable \( z \) is set to 4, while the number of channels in the condition variable output from the conditional encoder is set to 16. During the diffusion phase, we used a training schedule of \( T=1000 \) time steps and used a linear noise scheduler ranging from \( 1 \times 10^{-4} \) to \( 2 \times 10^{-2} \).

The three stages of training were optimized using the Adam optimizer with learning rates of \( 1 \times 10^{-4} \), \( 5 \times 10^{-5} \), and \( 5 \times 10^{-5} \), respectively, over 1000 epochs, saving the checkpoint with the minimum validation loss. Due to memory constraints, the batch sizes for the three training phases were set to 1, 8, and 2, respectively.
\begin{figure}
    \centering
    \includegraphics[width=\linewidth]{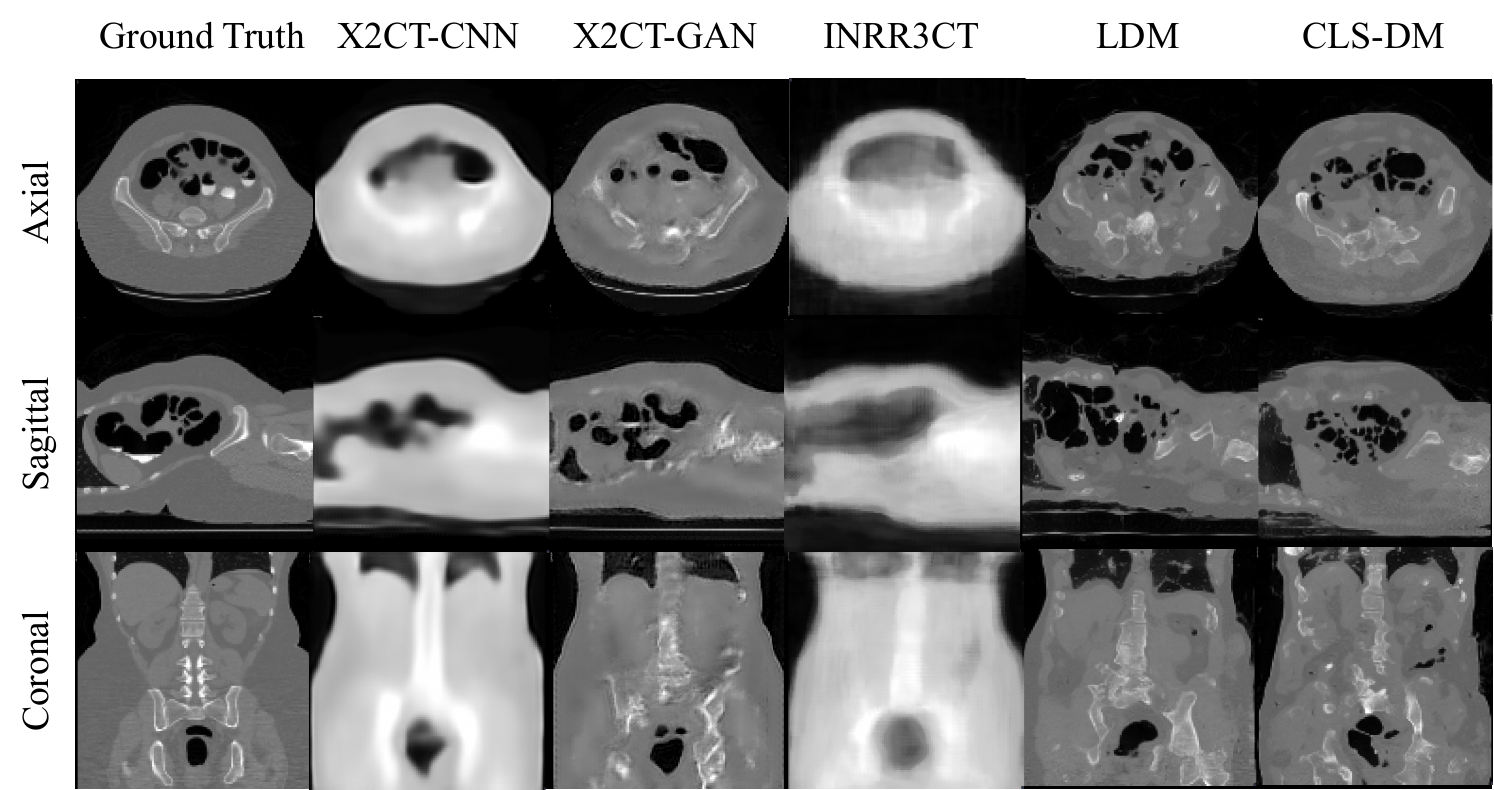}
    \caption{Visual comparison between the results of various models and the Ground Truth under the CTSpine1K dataset}
    \label{fig:ctspine}
\end{figure}

\subsubsection{Baselines}
We compared CLS-DM with four different baselines, including X2CT-GAN, X2CT-CNN, INRR3CT, and LDM. In particular, X2CT-CNN employs an Encoder-Decoder framework, while X2CT-GAN incorporates a discriminator to create a GAN framework based on X2CT-CNN. INRR3CT represents a network architecture that is grounded in CNNs and implicit neural representations (INRs). LDM serves as the backbone of CLS-DM, and further discussions regarding this can be found in the ablation studies section (see \ref{sec:ablation}). Additionally, we tested the 3D Diffusion approach, which directly implements diffusion on CT and X-ray images in voxel space. However, due to memory limitations, the results obtained were significantly inferior compared to other baselines, and thus are not discussed in this context.

\begin{figure}
    \centering
    \includegraphics[width=\linewidth]{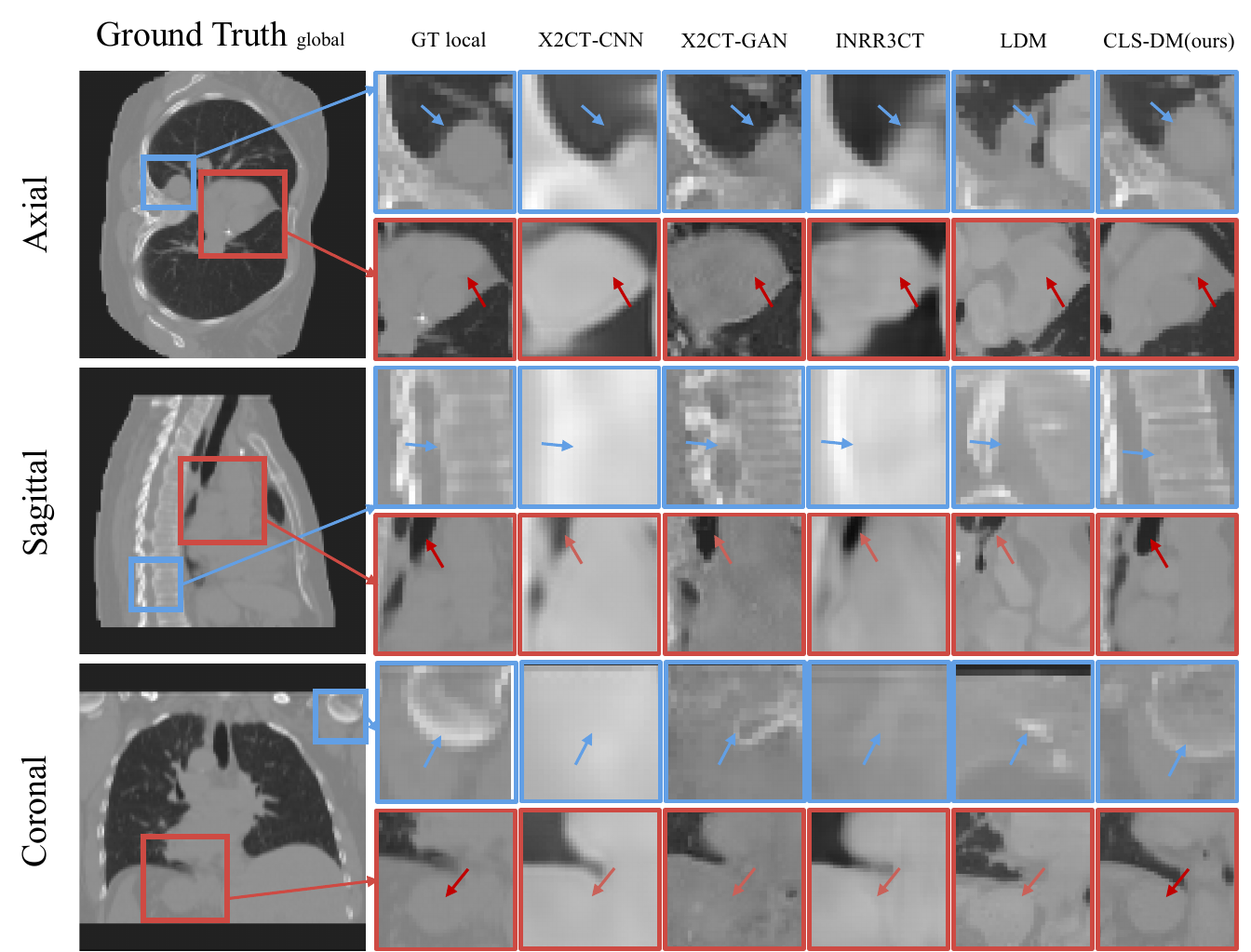}
    \caption{A detailed comparison of the results obtained from various models and the Ground Truth under the LIDC-IDRI dataset is presented. For each slice, we conduct a comparative analysis of specific regions designated by a 16×16 area (indicated by the blue box) and a 32×32 area (indicated by the red box). Corresponding arrows are utilized to highlight areas of particular interest.}
    \label{fig:lidc_detail}
\end{figure}

\subsection{Quantitative and Qualitative Results}
\subsubsection{Quantification}
We organized two datasets, with experimental results for each dataset utilizing frontal, lateral, and double-view configurations (see Table \ref{tab:comparison}). The CLS-DM model consistently outperformed the baseline models in terms of PSNR and SSIM metrics across both datasets. Notably, CLS-DM exhibited a slightly lower PSNR when tested with the 2-view and lateral perspectives on the LIDC-IDRI dataset compared to the X2CT-CNN model. This discrepancy can be attributed to the fact that PSNR is calculated based on mean squared error (MSE), and the optimization objective of X2CT-CNN specifically targets MSE, leading to somewhat blurred visual outputs (refer to the subsequent Qualification section). Overall, when averaging all metrics across the three perspective conditions in both datasets, CLS-DM surpassed the second-best results by nearly two points in both PSNR and SSIM.

\begin{figure}
    \centering
    \includegraphics[width=\linewidth]{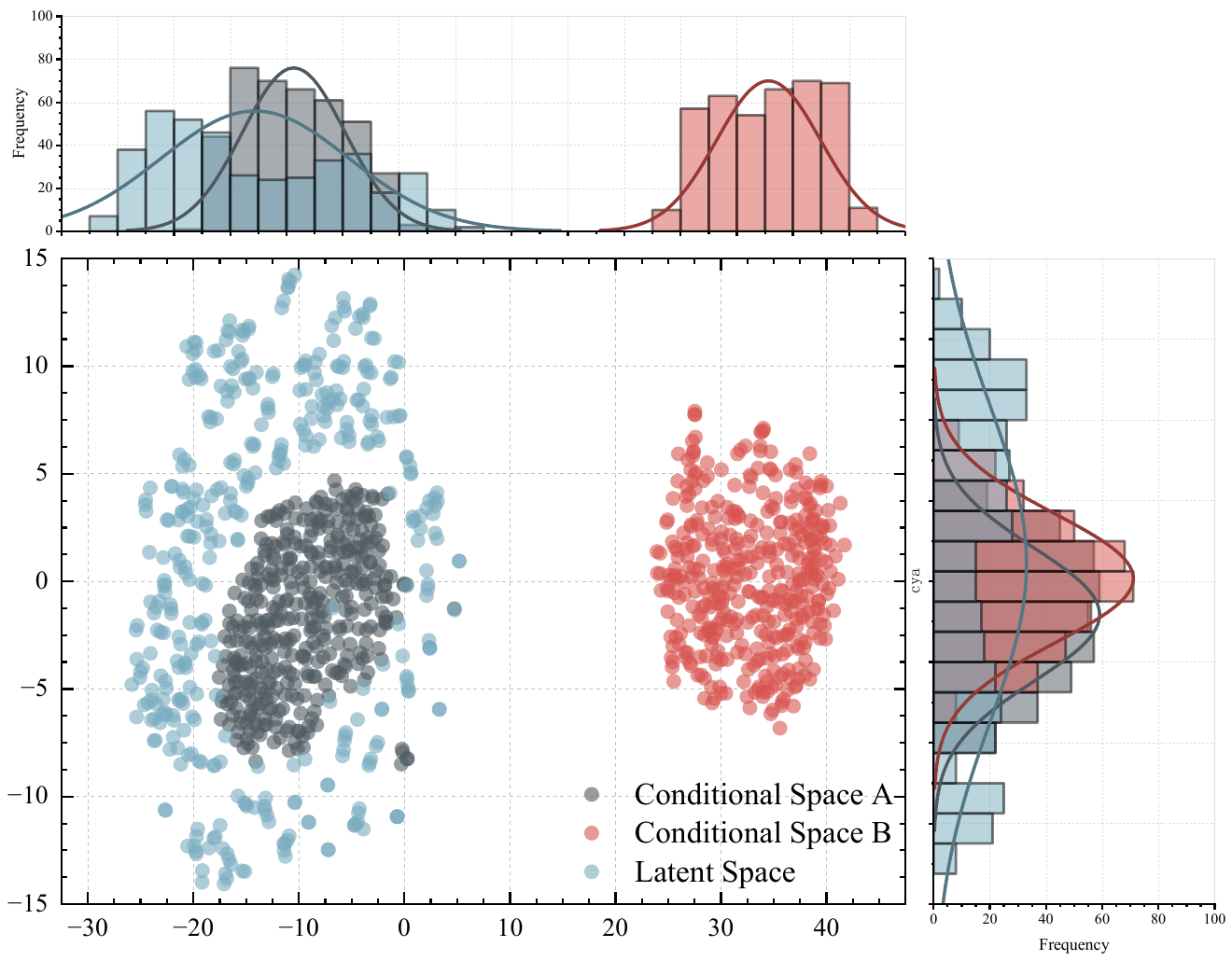}
    \caption{The results illustrate the distribution of X-ray feature sampling and latent space distributions before and after contrastive learning, as analyzed using t-SNE dimensionality reduction on the LIDC-IDRI dataset. "Condition Space A" denotes the distribution of X-ray features after contrastive learning, which acts as a conditioning variable in the diffusion process, while "Condition Space B" pertains to the distribution prior to the application of contrastive learning.}
    \label{fig:space}
\end{figure}

In addition, we conducted a quantitative assessment of the influence of contrastive learning guidance. We applied t-SNE for dimensionality reduction of the sampled points, resulting in the distribution visualized in Figure \ref{fig:space}. In the upper and right sections of the figure, we plotted the distribution functions for the three spaces post-dimensionality reduction. It is evident that contrastive learning explicitly aligns the distribution within the latent space without altering the inherent patterns of X-ray feature distributions.

\subsubsection{Qualification}
Figures \ref{fig:lidc} and \ref{fig:ctspine} present a comparative analysis of the results obtained from the various models against the ground truth for the same example in both the LIDC-IDRI dataset and the CTSpine1K dataset. It is evident that the models utilizing mean squared error (MSE) as their optimization objective, namely X2CT-CNN and INRR3CT, yield overly smooth and averaged results. In contrast, the outputs generated by the diffusion-based models, LDM and CLS-DM, exhibit significantly more detail than those produced by the GAN-backboned X2CT-GAN. Under the corrective influence of contrastive learning and autoregressive constraints, CLS-DM demonstrates greater accuracy in the results obtained under identical training conditions compared to LDM. Figure \ref{fig:lidc_detail} highlights detailed sections from the LIDC-IDRI dataset, showcasing that CLS-DM provides substantially more pronounced and accurate details across various regions in the three cross-sections.

\subsection{Ablation Studies}
\label{sec:ablation}
We conducted ablation experiments on the CLS-DM model utilizing the LIDC-IDRI dataset, with the results summarized in Table \ref{tab:ablation}. The experiments involved conditions without the contrastive learning process (denoted as w/o CLIP), which represents the use of a two-stage training approach incorporating an autoregressive constraint via a conditioning encoder during the diffusion process; a scenario where the contrastive learning phase lacks the autoregressive constraint (designated as w/o AR); and an instance where both contrastive learning and autoregressive constraints are excluded, resulting in the degradation to a backbone latent diffusion model (LDM). 

A comparative analysis of the results for “w/o AR,” “LDM,” and “CLS-DM” indicates that the training strategy employing an autoregressive constraint with a condition encoder, guided by contrastive learning, effectively enhances the model's expressiveness and reconstruction capability. Conversely, reliance solely on contrastive learning tends to compromise the feature representation ability of the condition encoder.
\begin{table}
\centering
\caption{The table presents the results of the ablation experiments. The designation "w/o CLIP" indicates the absence of the contrastive learning process; "w/o AR" signifies the exclusion of the autoregressive constraint; and "w/o CLIP, AR (LDM)" refers to the condition in which both are absent while employing the backbone LDM. The optimal values are marked in bold, whereas the suboptimal values are highlighted with underlining.}
\resizebox{0.48\textwidth}{!}{
\begin{tabular}{c|cc|cc|cc}
\toprule
Method/X-Type & \multicolumn{2}{c|}{Frontal} & \multicolumn{2}{c|}{Lateral} & \multicolumn{2}{c}{2-view} \\
\cline{2-7}
 Metrics& PSNR↑ & SSIM↑ & PSNR↑ & SSIM↑ & PSNR↑ & SSIM↑ \\
\hline
w/o CLIP & 22.42 & \underline{0.5919} & \textbf{25.17} & \textbf{0.6591} & \underline{25.37} & \underline{0.6701} \\

w/o AR & \underline{22.66} & 0.5083 & 23.53 & 0.5735 & 24.03 & 0.6012 \\

w/o CLIP,AR(LDM) & 20.65 & 0.5864 & 24.73 & 0.6473 & 24.94 & 0.6515 \\
\midrule
CLS-DM (ours) & \textbf{23.02} & \textbf{0.5931} & \underline{24.96} & \underline{0.6585} & \textbf{27.36} & \textbf{0.7058} \\
\bottomrule
\end{tabular}
}
\label{tab:ablation}
\end{table}

Furthermore, a comparison between “w/o CLIP” and “CLS-DM” reveals that the inclusion of contrastive learning significantly improves reconstruction performance when both Frontal and Lateral views of X-ray images are utilized. However, this enhancement is not evident when reconstructing using only a single Frontal or Lateral view. This phenomenon may be attributed to the simplicity of the condition space when a single X-ray is employed for reconstruction, which diminishes the observable effects of contrastive learning guidance. In contrast, the complex condition space arising from reconstruction using dual viewpoints underscores the critical importance of contrastive learning guidance. This finding highlights the potential of contrastive learning guiding multi-view reconstructions.

\section{Disscusion and Conclusions}
In this paper, we introduce CLS-DM, a latent diffusion model explicitly guided by contrastive learning to align the latent spaces between CT and X-ray modalities. The training of CLS-DM consists of three stages: the first stage compresses CT images from voxel space into latent space; the second stage employs contrastive learning guidance and autoregressive constraints to train the X-ray feature encoder, thereby aligning the latent spaces between the modalities; the third stage utilizes the feature encoder, completed in the previous stage, as an auxiliary condition to train the diffusion model in the latent space. Through extensive experimentation, CLS-DM has demonstrated its capacity to efficiently and accurately reconstruct CT images using sparse view X-ray data.

Our subsequent work will focus on leveraging the concept of feature pyramids to further constrain the training of the feature encoder through multi-stage and dimensional feature embeddings, and we aim to extend this work to other domains. However, this study is not without its limitations: while contrastive learning explicitly aligns the latent spaces between modalities, it somewhat compromises the feature representational capacity of the X-ray feature encoder. We will continue to refine CLS-DM with the objective of achieving more efficient and accurate reconstruction of CT images from sparse view data.

\FloatBarrier
\bibliographystyle{ACM-Reference-Format}
\bibliography{sample-base}

\end{document}